SURVEY

# Machine Learning for Healthcare-IoT Security: A Review and Risk Mitigation

**MIRZA AKHI KHATUN**[1,2], (Member, IEEE), **SANOBER FARHEEN MEMON**[1], (Member, IEEE),
**CIARÁN EISING**[1,2], (Senior Member, IEEE),
**AND LUBNA LUXMI DHIRANI**[1,2], (Senior Member, IEEE)
[1]Department Electronic and Computer Engineering, University of Limerick, Limerick, V94 T9PX Ireland
[2]SFI CRT Foundations in Data Science, University of Limerick, Limerick, V94 T9PX Ireland

Corresponding author: Mirza Akhi Khatun (Mirza.Akhi@ul.ie)

This work has emanated from research conducted with the financial support of Science Foundation Ireland (SFI) under Grant Number 18/CRT/6049.

**ABSTRACT** The Healthcare Internet-of-Things (H-IoT), commonly known as Digital Healthcare, is a data-driven infrastructure that highly relies on smart sensing devices (i.e., blood pressure monitors, temperature sensors, etc.) for faster response time, treatments, and diagnosis. However, with the evolving cyber threat landscape, IoT devices have become more vulnerable to the broader risk surface (e.g., risks associated with generative AI, 5G-IoT, etc.), which, if exploited, may lead to data breaches, unauthorized access, and lack of command and control and potential harm. This paper reviews the fundamentals of healthcare IoT, its privacy, and data security challenges associated with machine learning and H-IoT devices. The paper further emphasizes the importance of monitoring healthcare IoT layers such as perception, network, cloud, and application. Detecting and responding to anomalies involves various cyber-attacks and protocols such as Wi-Fi 6, Narrowband Internet of Things (NB-IoT), Bluetooth, ZigBee, LoRa, and 5G New Radio (5G NR). A robust authentication mechanism based on machine learning and deep learning techniques is required to protect and mitigate H-IoT devices from increasing cybersecurity vulnerabilities. Hence, in this review paper, security and privacy challenges and risk mitigation strategies for building resilience in H-IoT are explored and reported.

## I. INTRODUCTION

The Internet of Things (IoT) consists of interconnected physical devices exchanging data through sensors, software, and connectivity [1], [2]. The healthcare industry has undergone a significant transformation in recent years with advances in IoT, cloud, artificial intelligence (AI), and machine learning (ML). According to several experts, the expanding horizon of IoT is expected to improve healthcare. IoT can revolutionize healthcare globally by providing affordable healthcare [3], remote health monitoring [4], wellness management [5], and virtual rehabilitation [6]. Healthcare analytics can provide insight into disease and drug discovery while adding a new dimension [7].

The associate editor coordinating the review of this manuscript and approving it for publication was Claudio Agostino Ardagna.

The modern world requires more efficient and timely interventions to combat escalating health issues. While traditional healthcare systems are effective, these systems are often slow and inflexible [8]. The COVID-19 pandemic has fueled the need for remote and precision healthcare, and such objectives could only be achieved using emerging technologies. Embedding an IoT-enabled architecture in a healthcare ecosystem may facilitate the collecting and processing real-time data from sensors (i.e., body sensors strategically placed on or within a patient's body, aiding real-time data collection) [9]. Different sensors are used for different applications, such as motion, flow, and biomedical. However, the ones used for healthcare applications are typically body sensor networks (BSN) [10]. In a BSN, each sensor is connected within a group, forming a network; this configuration enhances data collection and efficiency by integrating IoT body sensors, such as brain waves, body temperature, and blood pressure sensors, as depicted in Figure 1. Moreover, the diagram also illustrates how data is stored and transmitted to the cloud via edge nodes/layers for data processing and storage. IoT devices transmit data in milliseconds and require high reliability, scalability, and end-to-end (E2E) latency connectivity networks [1]. Connecting the IoT devices with an edge-to-cloud environment is essential to facilitate the required/comprehensive health monitoring (security and reliability metrics).

As per [11], the need and usage of IoT devices will increase significantly; by 2025, there will be more than 41 billion IoT devices used worldwide, with the capacity to produce 78 zettabytes of data. Because of its usability, efficiency, and applications, these devices have the potential to further advance healthcare by reducing costs, improving patient well-being, and facilitating the efficient delivery of faster diagnostics, thereby improving medical services [12]. Furthermore, IoT can provide the migration of patients from traditional healthcare management methods to new cloud-based systems, among other vital innovations [13]. Employing Healthcare-IoT (H-IoT), remote monitoring of patients can save millions of lives and money, while other functions still have a crucial role across healthcare environments [14]. This type of remote monitoring of patients can assist in identifying health issues early on and making treatment plans more personalized. Similarly, IoT can benefit medical management, including medication and instrumental errors, and medication administration-allowing patients to take their medications at the right time and with the correct dosage [15]. Medication error prevention programs can improve patient outcomes and reduce healthcare costs. For example, H-IoT could significantly improve health outcomes and healthcare costs for residents of care homes who frequently manage multiple medical conditions and take various medications [16].

IoT healthcare frameworks utilize a variety of sophisticated sensors, including diagnostic, sensitive, and preventive sensors, for the implementation of healthcare systems [17]. Over the past few years, healthcare researchers have primarily focused on finding ways to monitor patients remotely and transmit health reports in real-time to physicians. Researchers have identified several major challenges related to health surveillance systems. The challenges include data privacy, interoperability issues, data quality issues, and limitations associated with real-time analysis [18], [19]. Patient monitoring concerns can be classified into two categories: static and dynamic monitoring. Smart hospitals use static monitoring systems to record patients' health status periodically. Moreover, as a static monitoring system, medical staff, including doctors and nurses, manually collect patients' vital signs during specific periods. The frequency of manual data collection within hospitals varies based on factors such as physicians' workload, working hours, patients' health conditions, and hospital leadership guidance [20].

In contrast to static patient monitoring systems, dynamic patient monitoring systems can be used at home, work, or in the hospital. According to Figure 1, healthcare systems consist of identifying, locating, sensing, and connecting. H-IoT components include emergency medical services, information technology, sensors, lab-on-a-chip technologies, wearable devices, connectivity devices, big data, and cloud computing [21].

Nowadays, smartphones enable rapid task completion by monitoring and collecting regular updates on patients' healthcare data [22]. For instance, an individual can promptly receive notifications when their heart rhythm changes when they wear a wristband linked to their smartphone [23]. Furthermore, IoT allows the healthcare system to monitor and track community resources more efficiently and reliably [24]. However, H-IoT has several risks, including privacy leakage during medical data uploading [25]. The risk of patient data disclosure may considerably discourage patients from sharing their medical information because confidentiality is at risk [26]. Maintaining confidentiality and credibility in healthcare interactions are fundamental factors for addressing these risks. Moreover, resolving such security concerns will facilitate seamless deployment and adoption of H-IoT.

Over the years, several review papers have been published on the secure H-IoT [27], [28], [29], [30], [31], [32], [33], [34], [35], [36], [37], [38]. There is, however, a noticeable gap in the existing review regarding the security of H-IoT, often failing to address all necessary aspects holistically. Most of the discussions focus heavily on the technical aspects of H-IoT devices, diving deep into their specs and applications and skipping over the essential issue of data security. During the discussion of machine learning, the focus is on its potential application in healthcare instead of exploring how it can enhance security protocols, such as detecting unusual data patterns that may indicate a cybersecurity threat. Security issues related to remote patient monitoring in H-IoT have not been discussed in existing reviews either. In [39], the authors analyze the Internet of Medical Things (IoMT) security controls in a sustainable context but do not address specific issues like remote patient monitoring or the use of AI in IoMT security. In [40], IoT and AI are explored, and a smart pill bottle case study is presented; however, detailed comparisons of these technologies appear lacking, and alternative cybersecurity strategies may not be fully considered. In [41], the authors reviewed the wireless body area network-based IoT healthcare systems but did not provide a detailed exploration of real-world implementations and case studies. Furthermore, although the discussion acknowledges security and privacy concerns, it does not provide in-depth solutions or strategies to address identified challenges within IoT healthcare systems. In [42], a review of the Internet of Healthcare Things (IoHT), covers its technologies, applications, and inherent challenges, particularly security and privacy concerns. Though it explores many issues, notably privacy and security, it modestly limits its explanations to practical solutions. In [43], the authors discuss the development of fifth-generation (5G) technology within the H-IoT space, including network slicing, security, and energy efficiency. Despite this, certain topics remain unresolved and require more exploration and innovation. These include data security, efficient resource management, and energy conservation. Healthcare services use large amounts of energy for heating, ventilation, and air conditioning (HVAC) [44]. HVAC systems can adjust their settings in real time using IoT sensors.

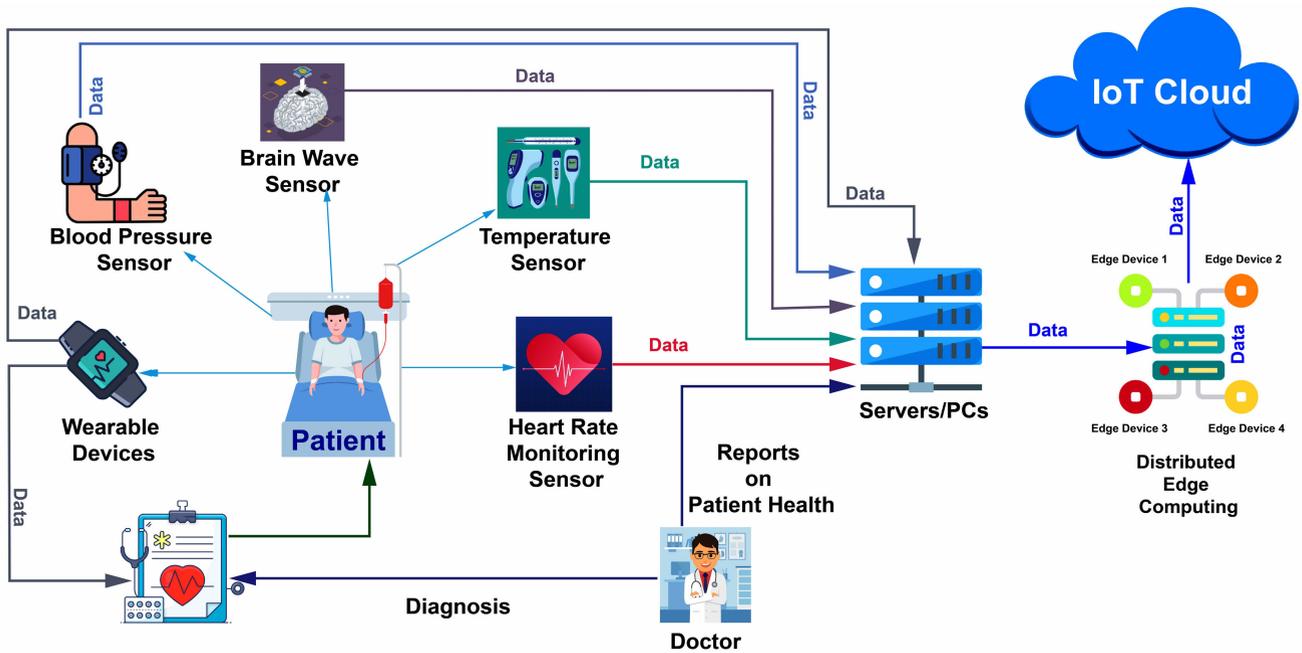

**FIGURE 1.** IoT based healthcare architecture.

In H-IoT research, secure data access presents a pivotal challenge, particularly regarding healthcare data regulations and strict policies that protect patient identity/confidentiality. Looking at the various ransomware attacks and data breaches [45] that healthcare sectors have suffered in the past, affecting millions of patients globally, demonstrates the urgency and need to mitigate these growing cyber risks. The COOJA simulator has reshaped This complex research area, allowing researchers to emulate H-IoT device behaviors without interacting directly with sensitive data. Data collection from H-IoT devices was not discussed in previous surveys. In addition, there was a limitation on how malicious data could be collected for anomaly detection by ML. This paper reviews the security aspects of H-IoT applications, pinpoints vulnerabilities, and uses ML to suggest solutions, providing an enhanced understanding of the security issues associated with H-IoT that incorporates both practical and theoretical considerations. In addition to highlighting the mitigation of H-IoT security challenges, this review also illustrates the use of COOJA simulator tools, which allows a unique perspective to be gained that makes this survey distinctive from those previously conducted.

This paper aims to review current research on AI and ML techniques that can enhance H-IoT security, build cyber resilience within the healthcare infrastructure, and enable it to detect, protect, and respond to novel cyber threats. The paper also covers a broad spectrum of cybersecurity issues from the European Union Agency for Cybersecurity (ENISA) 2030 cyber threat landscape foresight [46] perspective, including complexities associated with IoT layers. Furthermore, this review outlines promising avenues for future research and highlights the advancements in the field. The paper highlights the following key contributions:

- Reviews key aspects of cybersecurity, big data, e-health, and cloud computing in H-IoT.
- Discusses machine learning techniques such as anomaly detection, device classification, and critical use cases for security enhancement such as intrusion detection, authentication, and access control.
- Highlights the challenges and potential solutions related to H-IoT security, along with future research directions.

Figure 2 presents the structure of this survey paper as follows: section II explains 26 different types of cyber attacks carried out across the four main layers of the H-IoT architecture, section III describes the interaction between emerging technologies, such as machine learning and cloud computing, e-Health system and also discusses the risks associated to Healthcare-IoT, section IV emphasizes risk mitigation strategies, reviews the evolution of cybersecurity challenges in healthcare over the last five years, and provides a comparative analysis of H-IoT attack datasets, section V explores specific H-IoT layers, including data collection, applications with an emphasis on routing attacks, and network layers, as well as COOJA simulator, section VI focuses on feature extraction, intrusion and event detection, system reliability, and machine learning techniques to enhance cybersecurity, and convolutional neural network in the H-IoT, section VII consists of solutions designed for future H-IoT environments, including anomaly detection, security, and privacy, section VIII outlines the future of H-IoT technologies, software defined networks, challenges, and mitigation strategies and finally, section IX concludes the paper, exploring future directions in H-IoT security.

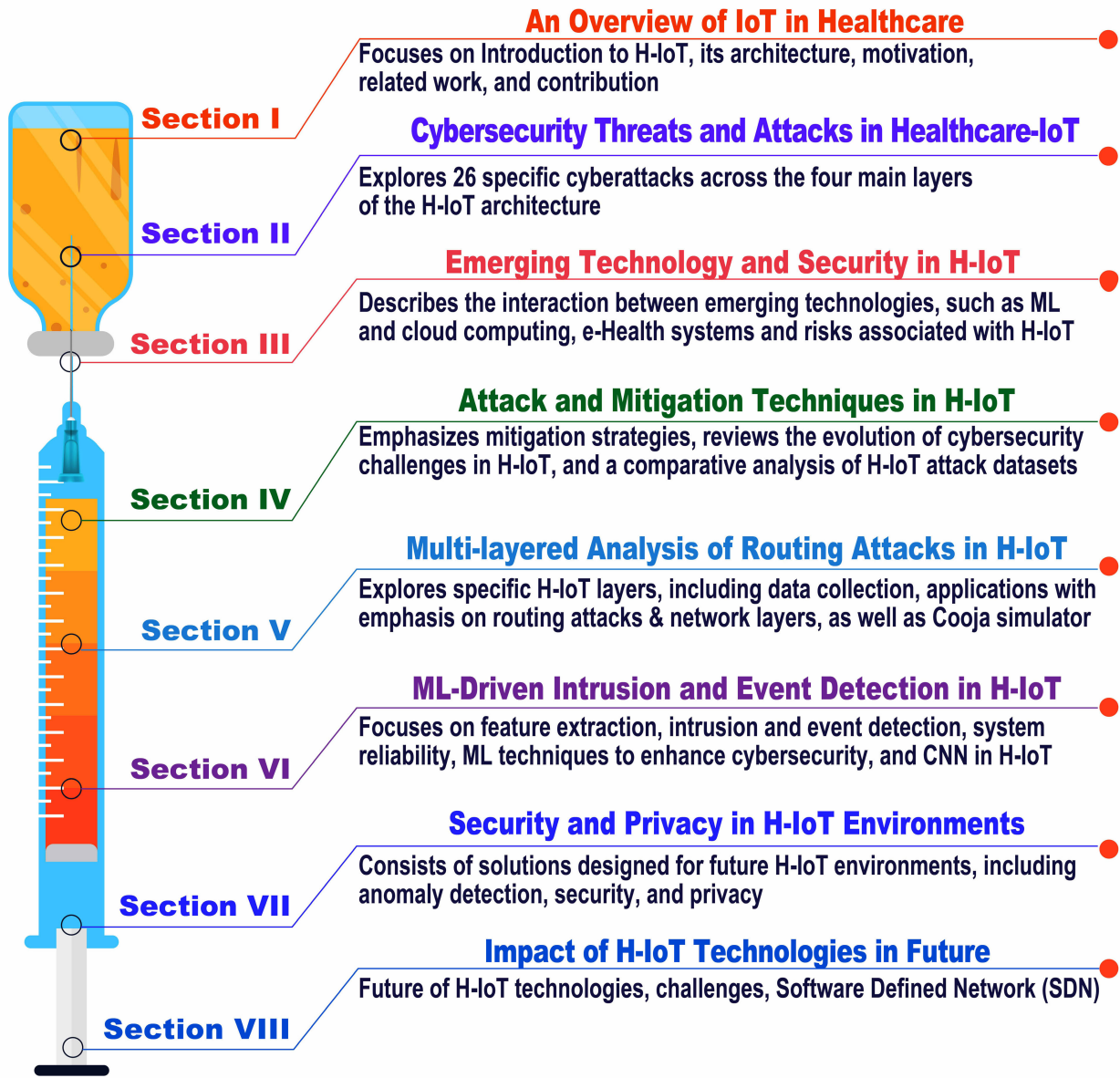

**FIGURE 2.** Structure and outline of the paper.

## II. CYBERSECURITY THREATS AND ATTACKS IN HEALTHCARE-IoT

Healthcare Internet of Things (H-IoT) requires a thorough understanding of the technical architecture, typically categorized into four layers. These layers include (I) the perception layer, (II) the network layer, (III) the cloud or processing layer, and (IV) the application layer [47]. Cybersecurity threats and attacks emerge at these layers, causing significant challenges (i.e., routing attacks, impersonating, tampering, or data transit attacks at the perception layer, denial of service and distributed denial of service (DoS/DDoS) or man-in-the-middle (MITM) attacks at the network layer, Cloud-malware injection or Brute-force attacks at the Cloud-IoT Layer and SQL injection or Cross-Site Scripting (XSS) attacks at the application layer) [48], [49], [50], [51], [52], [53], [54]. This section discusses 26 different cyberattacks and demonstrates their specific mechanisms and impact on the infrastructure. Figure 3 illustrates the details of healthcare IoT layers. It also provides a detailed explanation and insights into the most prevalent threats and attacks at different H-IoT layers.

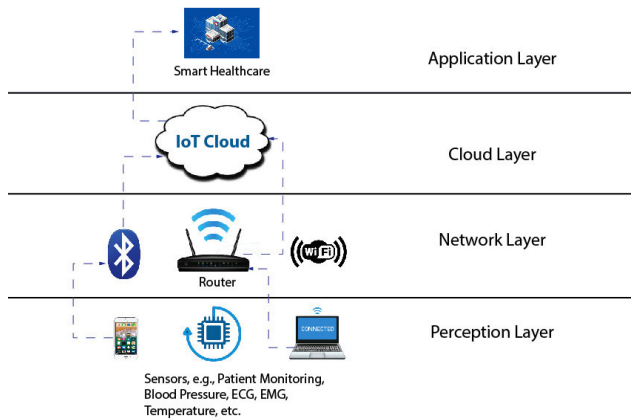

**FIGURE 3.** The layers of H-IoT.

### 1) THE PERCEPTION LAYER

This layer can also be termed the ''physical layer,'' or the ''sensor layer,'' as it refers to the physical devices [55], [56] that serve the purpose of sensing and collecting essential information about patients, such as their medical history. IoT healthcare frameworks enable the interconnection of key stakeholders such as doctors, nurses, technicians, pharmacists, and medical devices as part of the perception layer. A specialist can monitor the data the perception layer collects in real time over the Internet. There are multiple communication protocols (i.e., Radio Frequency Identification (RFID), Bluetooth Low Energy (BLE), Wireless Sensor Networks (WSNs), Zigbee, and IPv6 over Low-Power Wireless Personal Area Networks (6LoWPAN)) used for collecting and transmitting data from IoT nodes [57], [58], [59], [60]. This layer is susceptible to multiple attacks affecting the infrastructure in various ways, as depicted in Figure 4.

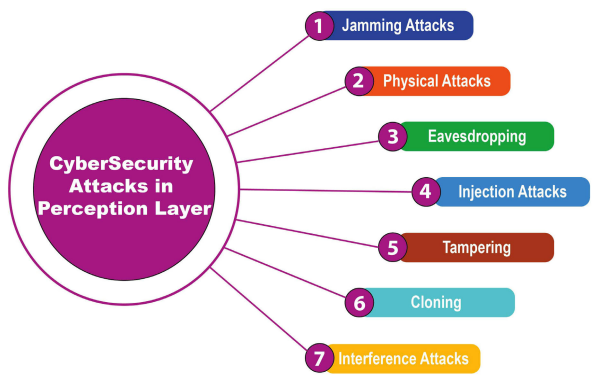

**FIGURE 4.** Cybersecurity attacks in perception layer.

Here, we summarise each threat in this layer briefly:
- **Physical Attacks:** These attacks focus on the IoT architecture's perception layer or physical layer consisting of hardware devices. To attack this layer, the malicious actor must remain closer to network infrastructure or gain unauthorized physical access to execute an attack [61]. These exploitations may facilitate the threat actor to tamper with implanted medical devices (i.e., Wi-Fi-enabled pacemakers and insulin pumps).
- **Eavesdropping:** IoT devices are vulnerable to eavesdropping if such attacks are executed successfully, which gives the malicious actor direct access to retrieve confidential information exchanged between the devices or radio frequency identification (RFID) tags and readers [62], [63], [64]. Eavesdropping attacks may have a variety of malicious intents, such as taking intellectual property, biometrics, and genome data for personal gain or using personally identifiable information/genetic data for espionage [65]. Such attacks have been carried out in the past [66].
- **Jamming:** The majority of wireless devices communicate with each other via radio frequency (RF) signals, which stronger signals can obstruct. In this scenario, a malicious actor may interrupt and block communications between sensors and receivers, leading to an absolute downtime and lack of availability [67], [68]. Jamming stands out as one of the most common attacks carried out at perception layer attack, alongside physical tampering, false data injection, and eavesdropping [69], [70], [71], [72], [73], [74], [75], [76]. The impact of jamming attacks in H-IoT can be catastrophic as it could disrupt on-going surgeries, medical diagnostics, access to online-systems. In short, it can lead the healthcare facility to an absolute downtime, affecting human lives. These types of cyber-attacks have an escalated impact on an infrastructure's operational and financial aspects and the ethical/social dimensions.
- **RFID Cloning:** These attacks have emerged as a major security concern at the IoT's perception layer as they camouflage themselves in different forms, such as RFID cloning and tag cloning. RFID cloning involves replicating and deceiving readers with duplicate tags that simulate the original to gain unauthorized access to information [77], [78]. The concept of tag cloning, however, encompasses duplicating various tag-based identification systems beyond RFID [79]. These two forms of hacking endanger the system's integrity and increase risks related to bio-hacking [80].
- **Injection Attacks:** These attacks can occur on various IoT layers; however, at the perception layer, it is carried out by injecting malicious code and modifying the firmware of an IoT device. IoT infrastructures face the greatest challenge when a single compromised/malicious node can spread across and infect the entire network, causing complete operational disruptions [81]. As per the European Cyber Resilience Act, digital and critical infrastructures must have the ability to detect, protect, respond, mitigate cyber threats, and maintain operational resilience simultaneously [82], [83].
- **Interference:** An intruder can disrupt network communication by interrupting network traffic and constantly broadcasting radio waves [84], for spreading false information. As per the European Network and Information Security Directive (NIS2-D), the healthcare infrastructures fall under the mandatory infrastructures to comply with the high common level of cyber security across the European jurisdiction [85]. These types of attacks can potentially spread misinformation and catalyze panic situations.

- **Tampering:** Attackers who manipulate nodes' memory to alter their functionality are called node tamperers. Furthermore, attackers may physically tamper with a device by turning it on and off, restarting it, stealing its key code, and manipulating the data [49], affecting the data integrity of the environment. Due to the misuse of generative AI, adversaries have been orchestrating and executing such attacks more frequently. As per the European General Data Protection Regulation (GDPR) [86], if any sensitive, personally identifiable information (PII) or healthcare data is breached in terms of integrity (data altered, amended, and tampered.), if so, it is considered as a data breach. An individual can suffer catastrophic consequences if their personal information is misused.

Defenses against perception-layer cyber-attacks in IoT are discussed in [87]. However, the approaches may not effectively mitigate the adversarial and escalating cyber threats due to generative AI.

### 2) THE NETWORK LAYER

Data packets are received and processed through this layer by the perception layer. Afterward, the layer transmits the made-trust data to the cloud layer immediately above it. The cloud layer's role is to store and share/build trust mechanisms between the smart devices [88], [89]. The network layer can be enabled with different wireless networks (Wi-Fi 6, 5G, Bluetooth, NB-IoT, and LTE) [90]. As part of the IEEE wireless communication standards, such as 802.11a, 802.11g, 802.11n (Wi-Fi 4), 802.11ac (Wi-Fi 5), 802.11ax (Wi-Fi 6), and 802.11p (for vehicular communications), orthogonal frequency division multiplexing (OFDM) is employed for the effective transmission of data across multiple frequency bands [91]. In the case of IoT devices, high-speed Wi-Fi has recently been recognized as an attractive option due to its compatibility with existing infrastructure [92]. The IEEE 802.11ax, a sixth-generation protocol popularly known as High Efficiency (HE), reports a 30% improvement in throughput over the older IEEE 802.11ac protocol known as very high throughput (VHT) [93]. A 5G-enabled energy-efficient routing protocol (EERP) that uses Wi-Fi 6 makes it easier for medical devices to communicate with one another, while multi-user, multiple input, multiple outputs (MU-MIMO) allows parallel communications [94]. A data-centric protocol prioritizes the transmission of critical health information, and content-centric networking (CCN) improves the efficiency of the distribution of healthcare information between IoT devices [95].

The vulnerability to DDoS attacks has increased dramatically with the advent of 5G. A distributed network architecture with multiple remote edge computing sites is essential for achieving low latency, which is critical for vehicle-to-vehicle communication. Additionally, the dense network of connected devices, open-source applications, and additional access points contribute to increasing "attack surfaces". Furthermore, higher wireless speeds dramatically increase the impact of DDoS attacks on the network since each device can transmit an order of magnitude higher attack throughput to the network (at MECs and the core). As mentioned, this effect is amplified by the plurality of IoT devices, which is expected in 5G, due to the vulnerability of IoT devices to cyberattacks [96]. Figure 5 illustrates seven attacks that are most encountered at this layer. The following discussion highlights key attacks in the network layer of the IoT:

- **Dos/DDoS Attacks:** In smart ecosystems, denial of service and distributed denial of service attacks are prevalent [97]. A hacker attempts to consume legitimate network resources or bandwidth during a DoS attack. This attack is referred to as a DDoS when it originates from multiple compromised nodes [98] and falls into different categories such as traffic/fragmentation attack, bandwidth attack, and application attack) [99]. Wireless and wired networks are critical to DDoS attacks, which target the network layer [100]. Flooding the Internet Control Message Protocol (ICMP) and User Datagram Protocol (UDP) with excessive data leading to delays, resource consumption, potential system crashes, and ultimately disrupting services are common methods for executing these attacks [101], [102]. These attacks would lead to absolute downtime and operational disruption.
- **Routing Attacks:** Throughout the Internet of Things, IPv6 is widely used, especially in wireless sensor networks. IPv6-centric WSNs are particularly vulnerable to routing attacks. Furthermore, WSN sensors are frequently constrained by memory limitations, narrow bandwidths, and energy consumption [48]. These attacks are generally carried out at the Internet Service Provider level, and Information and Communication

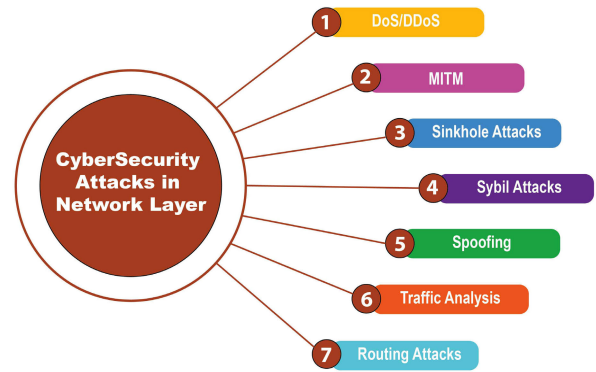

**FIGURE 5.** Cybersecurity attacks in network layer.

Technology (ICT) service providers must be NIS2-D compliant [103]. Detailed information about routing attacks can be found in section V-C.

- **Traffic Analysis:** Network traffic is analyzed for detecting and monitoring malicious behavior and anomalies. Depending on the network traffic, routers manage data packets. Each router is equipped with a certain amount of capacity. Clogging or unusually high traffic may indicate traffic analysis or even a denial-of-service attack [104]. Malicious actors sometimes observe the traffic for guessing passwords by analyzing the packets sent during each keystroke and assessing the duration between them. By using such a pattern, they can reconstruct the users' passwords.
- **Spoofing Attacks:** Several types of spoofing are possible, such as email, uniform resource locator (URL), and frame spoofing. However, MAC or IP address spoofing is most widespread [105]. IoT uses the MAC address to authenticate wireless networks at the data link layer. Spoofing attacks occur when malicious entities imitate legitimate users' MAC addresses to gain network access illegally, affecting the data confidentiality and integrity of the environment [106]. Lack of data confidentiality means unauthorized/malicious actors may know the contents of data, whereas lack of integrity means the data may have been amended, altered, tampered or discarded. As per GDPR, a lack of security (Confidentiality, Integrity, and availability) metrics and controls is considered a GDPR breach.
- **Sybil Attacks:** IoT devices are vulnerable to the Sybil attack [107], [108]. In this attack, legitimate nodes are impersonated by malware to redirect traffic towards malicious ones. An individual node can appear in multiple locations simultaneously or impersonate several others [109]. During this attack, a malicious node claims to have multiple identities. Due to their ability to control the flow of information within a network, these types of attacks affect data integrity and resource allocation.
- **Sinkhole Attacks:** In sinkhole attacks, the attacker redirected or discarded traffic, preventing the base station from receiving full data transmission. This attack spreads misleading routing details from one node to another, resulting in energy drain and reduced network durability. Energy depletion is particularly damaging to wireless sensor networks [110]. These types of attacks affect the integrity, availability, and reliability of data in a network.
- **Man-in-the-Middle (MITM) Attacks:** Man-in-the-middle attacks involve intercepting communications between two devices by an attacker [51]. The main target of MITM attacks is the confidential details of users [53]. Malicious actors exploit existing or new vulnerabilities in IoT systems to carry out this attack. An example would be a temperature reading of a sensor

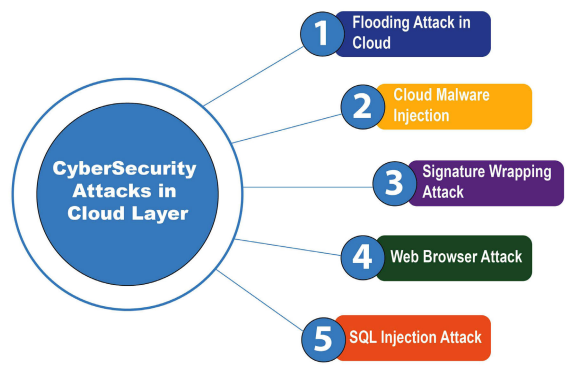

**FIGURE 6.** Cybersecurity attacks in cloud layer.

or malfunctioning of a device enabling hackers to steal sensitive information [111].

### 3) THE CLOUD LAYER

The cloud allows for convenient and secure backup and preservation of confidential health information, and sharing between authorized parties (such as doctors, insurance providers, medical staff, and pharmacies) is convenient. Similarly, independent and public healthcare providers can maintain confidential data online and share it with trusted colleagues to improve the quality of treatment. These protocols include transmission control protocol (TCP), user datagram protocol (UDP) [112], and protocols for data analysis, prediction, and machine learning [113]. Figure 6 depicts several attacks related to cloud layers. These specific vulnerabilities are explained in the following sections:

- **Flooding Attacks:** Flooding attacks are a type of DDoS attack that disrupt services by overloading the servers with enormous traffic from compromised computers. During these attacks, excessive messages are sent to servers, which causes legitimate users to be denied access to the Internet [114]. In the cloud environment, attackers often rely on sophisticated methods to exploit its inherent scalability and flexibility [115].
- **Web Browser Attacks:** Throughout the modern digital era, web browsers have become integral tools that allow users to access various online services and connect to the vast internet. Nowadays, browsers are a crucial component of almost every computer. Nevertheless, browsers are not immune from vulnerabilities [116], [117]. As a result of these vulnerabilities, attackers often gain access to a victim's computer, steal PII, corrupt files, or use the hacked machine to attack others, potentially becoming a part of a botnet [118].
- **Signature Wrapping Attacks:** Cloud infrastructure attacks may provide attackers with root-level access to systems without targeting the cloud environment directly [119]. The authentication systems of cloud interfaces are vulnerable to advanced cross-site scripting methods and signature wrapping [120]. In addition to manipulating virtual machine operations and password resets, signature-wrapping attacks are also known as XML signature-wrapping attacks [121]. In contrast, attackers can steal login credentials by exploiting the XSS vulnerability [122].

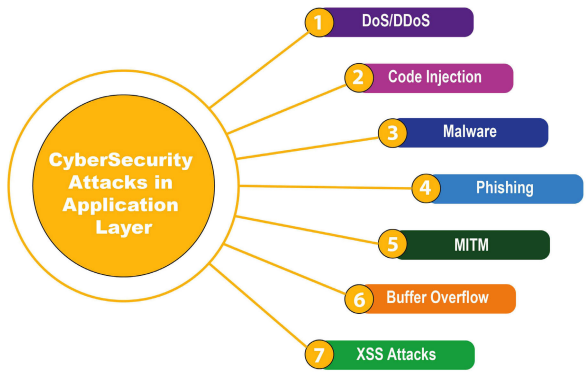

**FIGURE 7.** Cybersecurity attacks in application layer.

- **Cloud Malware Injection Attacks:** Attackers perform malware injections by inserting malicious code or services into a network, appearing as if these components were legitimate components already present [54]. As a result of this attack, users are often deceived into downloading software or opening malicious emails. This type of attack is sometimes driven by downloading or meta-data spoofing. An attacker may trick users into downloading harmful software by creating a deceptive environment. In the aftermath of such an attack, users may have limited access to their systems, while attackers may be able to execute malicious activities in the cloud and gain unauthorized control [123].
- **Structured Query Language (SQL) Injection Attacks:** Cloud computing often exposes web services to various security threats due to their inherent Internet accessibility. SQL injection is one of the most widespread risks [52]. By manipulating SQL queries, SQL injection attacks can achieve malicious objectives. Consequently, sensitive data may be modified unauthorized, confidential details may be accessed, or the server may crash due to such manipulation [124].

4) APPLICATION LAYER

Within the IoT architecture, the application layer resides at the top level. This layer provides the user interface, which includes services such as healthcare, smart homes, and connected cars [125], [126]. It recognizes spam and filters out the malicious content [127].

Figure 7 illustrates IoT applications' primary security attacks. The following sections delve into the application layer vulnerabilities:

- **DoS/DDoS Attacks:** DDoS attacks are carried out by a group of compromised computers operating from multiple locations that flood a particular target with traffic. In this attack, the primary goal is to overwhelm and render unavailable a server, website, or online service [128], [129]. In a DDoS attack, IoT devices such as smart appliances, healthcare monitors, and industrial sensors can be rendered useless [130]. In addition to affecting their immediate functionality, disruptions can jeopardize critical operations they oversee, like monitoring patient health or maintaining optimal building conditions [131].
- **Phishing Attacks:** Phishers can use compromised devices such as smartphones, appliances, and smart cars to conduct phishing attacks at the application layer. Through this method, attackers impersonate legitimate devices and send messages that appear to be authentic. A malicious attacker can then manipulate or obtain credentials to authenticate or identify, which can be used for criminal purposes [132]. For example, the Irish HSE Conti Cyber attack [133], which was initially a phishing attack, soon escalated to a ransomware attack after an employee accidentally clicked on the malicious email, disrupting the healthcare facility operations and IT outages across the country for days. However, it took four months for the Irish HSE to recover from the aftermath completely. This is why it is mandatory to build cyber resilience within such infrastructures.
- **Buffer Overflow Attacks:** Buffers hold data while it is transferred from one location to another. In buffer overflows, data exceeds the buffer's storage capacity. As a result, the application may crash, memory access errors may occur, and results may be incorrect. Memory overwrite vulnerabilities can be exploited by attackers. It can affect execution paths, leak confidential information, and corrupt files [134]. Legacy systems are most vulnerable to these attacks as they have limited memory [1].
- **Malware:** A malware attack attempts to use a malicious program to commit an offense with IoT applications, and recently, many malicious programs have been released to attack IoT devices, including rootkits, spyware, and adware [135], [136]. In [1] and [137], the authors outline different types of malware and their impact on the national security (i.e., Red October), social (transportation, communications, energy, and water sectors), financial (BaFin) [138] and economic domains (manufacturing, fintech) affecting human lives [139].
- **Cross-site Scripting (XSS) Attacks:** In cross-site scripting attacks, harmful code is embedded into an authentic and trusted website [140]. An attacker can alter the content of an application through this potentially dangerous intrusion [141]. Due to the inability of the targeted browser to distinguish between genuine and malicious code, the infected code is executed. Consequently, this malicious code can access cookies, session identifiers, or other confidential information. In addition, the attacker can control the device directly, sending users to malicious websites or causing direct damage to the device.

- **Unauthorised Scripts Attacks:** These attacks occur when unwanted or malicious scripts are executed without the user's consent. In contrast to XSS attacks, which target the user's browser, unauthorized scripts can run anywhere in the system or application. Data breaches, system malfunctions, or other vulnerabilities may result from this [142].
- **Code Injection Attacks:** Code injection attacks that use SQL injection break the data-code isolation rule by inserting malicious SQL codes into input fields [50]. Attackers can embed these malicious SQL commands in web forms, URLs, or page requests. Without proper filters, a web application may process these harmful commands incorrectly. As a result, unwanted access to the database can occur [143].

## III. EMERGING TECHNOLOGY AND SECURITY IN H-IoT

Technological advancements, including the IoT, machine learning, and cloud computing, are gradually reshaping healthcare. This development creates innovative healthcare solutions but also challenges, particularly when it comes to data security. AI and machine learning have made great progress in cyber security domains such as intrusion detection and malware mitigation [144]. This section explores the conjunction of machine learning and cloud technologies, investigating emerging trends and attendant security considerations within e-health applications to establish a secure and innovative framework for H-IoT implementations. This section further explores the connections between machine learning and cloud technologies within e-health, navigating through emerging trends and assessing related security concerns. Deep learning (DL) and machine learning are increasingly integral to mitigating cyber threats; however, how these technologies impact e-health is a crucial issue [145]. This section aims to provide a secure and innovative foundation for H-IoT implementations.

### A. MACHINE LEARNING WITH CLOUD COMPUTING

Researchers all over the world have applied machine learning to a variety of applications and domains [146]. Recently, ML has drawn the attention of H-IoT researchers [147]. In the context of H-IoT, machine learning is beneficial for remote monitoring and real-time treatment of diseases [19], [148]. ML algorithms such as Support Vector Machines (SVMs), decision trees, random forests, and Artificial Neural Networks (ANNs) can analyze huge volumes of medical data collected by healthcare-related smart devices, including vital signs and medical histories [149]. In this process, ML techniques are applied to analyze massive datasets to find patterns and generate insights that may assist clinical decisions, improve patient outcomes, and reduce healthcare expenditures. Cloud computing, however, provides computing power and storage resources for H-IoT to support machine learning based on big data collected from IoT devices [150]. ML algorithms utilize the cloud to process and analyze data and provide a secure and scalable computing environment [151].

Desai et al. [152] developed a Health-Cloud platform using machine learning and cloud computing to track patients suffering from heart-related diseases. Moreover, the researchers built a live data analysis iOS App using Google Cloud Firebase.

Abdelaziz et al. [153] investigate the use of IoT and cloud computing in healthcare to predict chronic kidney illness in a city of the future. IoT devices transmit chronic kidney disease (CKD) data to cloud storage, improving prediction accuracy. The hybrid model, which combines linear regression and neural networks, predicts CKD with 97.8% accuracy. Cloud IoT offers tremendous potential in healthcare services, benefiting patients and smart city stakeholders.

### B. ELECTRONIC-HEALTH (E-HEALTH) SYSTEMS

The exchange of health records between doctors and patients is now conducted largely electronically, using technology to transfer patient data efficiently and facilitate communications between doctors and their patients. As illustrated in Figure 8, e-health application systems can solve several problems associated with traditional healthcare systems, including online appointments, medical evidence, information technology, communication, e-prescribing, medical history, reminders, payment management, and lab analysis. e-healthcare services can enhance patient data and health information management [154]. As a result, patients can access their medical records online, receive remote consultations, and use mobile health applications to manage their health, which increases accessibility and empowers patients.

Maksimović and Vujović [155] describe the gradual adoption of e-health platforms primarily due to infrastructure and political restrictions. Despite these challenges, e-health and IoT convergence is progressing. However, there are hurdles like consistency, security, and interoperability in IoT integration. Privacy concerns and regulations further complicate the adoption of large-scale technologies. The paper highlights the influence of IoT in e-health and outlines these crucial challenges, emphasizing the importance of overcoming barriers to implementation. Zhang et al. [156] explore how 5G technology can revolutionize e-health. While 5G promises reliable access to e-health, current efforts are insufficient. This paper also discusses technological aspects, practical use cases, research trends, and challenges in advancing 5G e-health.

### C. BIG DATA IN HEALTHCARE-IoT

IoT networks generate enormous amounts of data at every moment. Therefore, manipulating so many datasets requires considerable technical expertise. IoT architecture and machine learning techniques boost big data capacity, and advanced deep learning models are essential [157]. Smart healthcare systems have dynamically transformed mathematical modeling with data collection using big data analytics. Besides collecting and storing data, big data is also about empowering machines to think and act like humans to simplify complex tasks [158], [159]. Wearable sensors continuously collect data, such as sleep patterns, exercise levels, walking distances, heart rates, etc. The latest IoT sensors can also monitor heart rate, blood sugar levels, and pulse. A major benefit of big data for the medical industry is that it could reduce the costs of accommodations, travel time, and transportation delays. In addition, big data analytics have improved healthcare facilities and enabled patients to recover at home [160]. H-IoT devices can also decrease medical expenses by reducing personnel and transport costs with big data [161], [162].

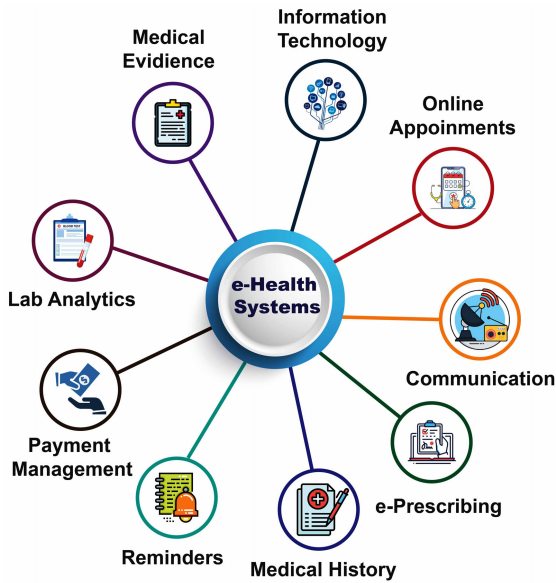

FIGURE 8. e-Health systems.

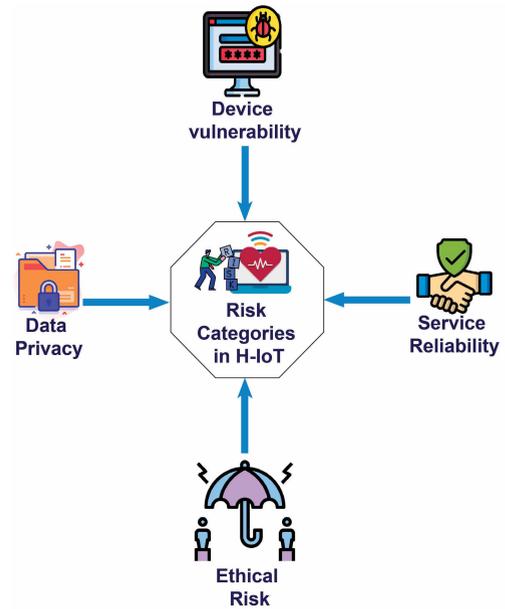

FIGURE 9. Risk categories in H-IoT.

Asri and Jarir [163] developed a real-time disease prediction platform that enhances patient care and reduces healthcare costs with a Big Data platform. It helps users make better decisions in real-time, improving health and safety. Heart attacks, obesity, and miscarriages can be detected using this approach. Long-term treatment and hospitalization costs can be reduced by early detection and accurate prediction of health problems.

Sasubilli and Kumar [164] utilize machine learning and big data to analyze the use of electronic health record applications in neuro- and cardiac-related fields. Integrating machine learning approaches and big data frameworks in healthcare architecture enhances data management while improving patient care. Advanced technologies enable more precise diagnostics, personalized treatment plans, and better outcomes, elevating patient care standards.

### D. RISK CATEGORIES IN HEALTHCARE-IoT

H-IoT brings multiple risks associated with medical devices integrated with advanced technologies. Data privacy, device vulnerability, service reliability, and ethical issues are all included in these risks. Both security and user trust depend on addressing these issues.

As depicted in Figure 9, these risk categories apply to the H-IoT industry. The following sections discuss each risk category and its impact within the context of H-IoT:

- **Data Privacy Risk:** During the COVID-19 pandemic, several countries reduced their data privacy guidelines as part of their emergency response [165]. Per the EU General Data Protection Regulation, implementing data security, governance, risk, and control metrics has been mandatory since the regulation was enacted in 2018; the regulation also enabled and improved the cyber security posture of organizations working across Europe in comparison to other countries/jurisdictions, and this is why Europe suffered lesser data breaches in comparison to the rest of the work. In the context of digitally enabled precision healthcare 5.0, securing systems and enforcing strong data controls are essential to mitigate risks associated with emerging technologies [166]; this would mean enabling end-to-end data (data-in-use, data-in-transit, and data-in-store) security, encrypting, anonymizing and pseudonymizing it, to mitigate the risks of data breach. Whenever adversaries can steal data that has been encrypted, anonymized, and pseudonymized, the data will have no value to the hackers [137].
- **Device Vulnerability Risk:** IoT has many security vulnerabilities, especially in the H-IoT space, because these devices would be manufactured by different suppliers who would not have considered security by design and privacy-by-design metrics, leading to unencrypted network services, inadequate password protocols, and user interface credentials. These issues have been identified as serious security lapses in 70% of H-IoT devices [167]. In addition, 90% of these devices collect personal data [168], illustrating the complexity and diversity of this issue. Healthcare sectors worldwide are particularly vulnerable to cyber threats due to their diverse operating environments and intricate regulatory frameworks. However, ENISA has passed new cybersecurity standards and regulations to combat supply chain, digital services, digital market, and third-party risks [137].

- **Service Reliability Risk:** In H-IoT, service reliability risks refer to service interruptions or failures, which can cause critical delays or data loss. This is a major concern for emergency response and ongoing health monitoring. H-IoT services require effective interference mitigation and optimization of 5G networks. To fully benefit from 5G and IoT, these steps will enhance reliability and efficiency [169]. With 5G and Beyond 5G (B5G) deployment, service interruptions, data loss, and seamless communications between multiple devices will be prevented in H-IoT or the Internet of Medical Things [170]. The ability to transmit real-time, accurate patient data to medical personnel through 5G and B5G is especially crucial in medical emergencies [171].
- **Ethical Risk:** A potential ethical risk associated with IoT devices is that of unethical actions [172]. There is an example from the automotive industry that serves as a cautionary tale, even though it isn't directly related to H-IoT. In violation of the Clean Air Act, Volkswagen developed and installed software to manipulate diesel emissions tests [173]. Reputational and financial damages resulted from this breach of ethics. When it comes to H-IoT, where patients' health and well-being are at risk, cyber ethical breaches such as harm to privacy, harm to property, and misuse of technical resources [137] can be detrimental [174].

## IV. ATTACK AND MITIGATION TECHNIQUES IN H-IoT

With cutting-edge technologies in the healthcare industry, the security and privacy risks have drastically increased [137]. As per [175], in 2022, more than 200 cyber attacks were carried out across healthcare organizations in the world daily by malicious actors, impacting millions of people worldwide. The number of cyber attacks on H-IoT increased by an alarming 74% in 2023 [176]. Demystifying cyber threats and attack scenarios is crucial to preventing these escalating risks. Therefore, this section explores the cybersecurity landscape of H-IoT. Table 1 surveys cybersecurity challenges within the healthcare sector over the last five years. Furthermore, this section discusses attack datasets for detecting and mitigating anomalies in H-IoT environments and threats, potential impacts, and mitigation strategies.

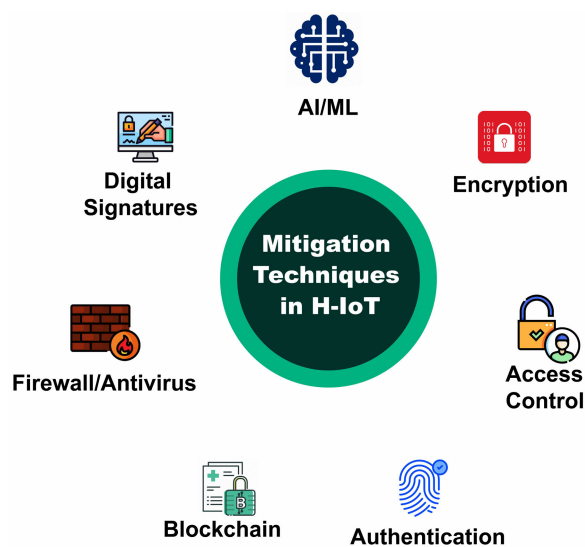

**FIGURE 10.** Mitigation techniques in H-IoT.

### A. MITIGATION TECHNIQUES

This section presents findings on mitigation strategies employed in prior research to mitigate cyber security threats associated with IoT in healthcare. Figure 10 illustrates the mitigation techniques for access control, encryption, authentication, AI/ML, blockchain, digital signatures, and firewalls/antivirus. H-IoT solutions also integrate modern AI and deep learning technologies. For instance, Al-Garadi et al. [177] described that machine learning models have been incorporated into security solutions to identify vulnerabilities and attack surfaces. Qiu et al. [178] explore the integration of healthcare and smart cities, enhancing health practices in countries worldwide. Blockchain technology addresses this integration's security concerns, ensuring the confidentiality of patient data. The authors discussed the benefits and drawbacks of each model used in the research at different IoT layers.

### B. COMPARATIVE REVIEW OF H-IOT ATTACK DATASETS

Datasets are crucial in identifying and mitigating anomalies in the ever-expanding world of the IoT in healthcare, industry, and general applications. H-IoT attack datasets are reviewed in this subsection, including simulations and real-world data from H-IoT sensors. In the constantly evolving area of H-IoT, these datasets serve as essential tools for research [179]. Similarly, machine learning applications require high-quality healthcare data, and exact labeling improves its accuracy [180], [181], [182]. Table 2 provides a detailed overview of datasets, attacks, mitigation techniques, benefits, and limitations.

## V. MULTI-LAYERED ANALYSIS OF ROUTING ATTACKS IN H-IoT

This section focuses on routing attacks in the H-IoT domain, reviewing key components sequentially. The first step is to discuss the vital process of collecting health-related data in the layer for medical data collection. After collecting raw medical data, the medical application layer transforms it into meaningful healthcare services. Next, this section discusses the role and vulnerabilities of the routing and network layers in the context of routing attacks. The next step is emphasizing that the medical application layer is crucial to translating data into actionable insights. In addition, this section discusses the role of the COOJA simulator in collecting data and generating datasets in the H-IoT environment.

TABLE 1. Cybersecurity challenges in H-IoT.

| Key Publications | Area | Security and Privacy Issues |
|---|---|---|
| Garcia-Perez et al. [183] (2023) | H-IoT | The paper highlights the lack of cyber risk awareness, management, and security risks related to emerging technologies. |
| Cartwright et al. [184] (2023) | IoMT, COVID-19 | Security gaps in protected health information (PHI), IoMT challenges, COVID-19 amplification, and funding shortages. |
| Javaid et al. [185] (2023) | H-IoT | Data security risks identified in the paper include ransomware attacks, unauthorized access to patient information, device vulnerabilities, and healthcare data source complexity. |
| Kumar et al. [186] (2023) | H-IoT | H-IoT devices face various security threats and scalability and latency challenges. |
| Sharma et al. [187] (2023) | DL, Cyber-attacks, Intrusion Detection System (IDS), Healthcare | A Deep Learning-based IDS detecting cyberattacks achieves 84% accuracy, increasing to 91% with Generative Adversarial Networks(GANs) balanced data. |
| Khatkar et al. [188] (2023) | DL, Intrusion Detection, Healthcare | For intrusion detection in healthcare, evaluating a DL-based Long Short-Term Memory (LSTM) algorithm against a Decision Tree, SVM, K-nearest neighbors (KNN). |
| Ksibi et al. [189] (2023) | IoMT, Cybersecurity, ML, Risk Assessment | Analyzes the security risks in the Internet of Medical Things (IoMT) using ML algorithms for anomaly detection and introducing of a new model of risk assessment. |
| Bouabida et al. [190] (2022) | Pandemic-driven, patient data risks | A significant concern raised by the COVID-19 crisis is the security and confidentiality of patient data. |
| Wang et al. [191] (2019) | Big Data, H-IoT | An Analysis of Big Data Analytics in healthcare is presented, including its advantages and challenges. |
| Coventry et al. [192] (2018) | Cybersecurity in Healthcare | Due to cybercrime risks, healthcare technology needs a holistic security program. |

### A. MEDICAL DATA COLLECTION LAYER
Healthcare data is collected by smart healthcare sensors, such as devices for tracking brain activity and blood pressure.

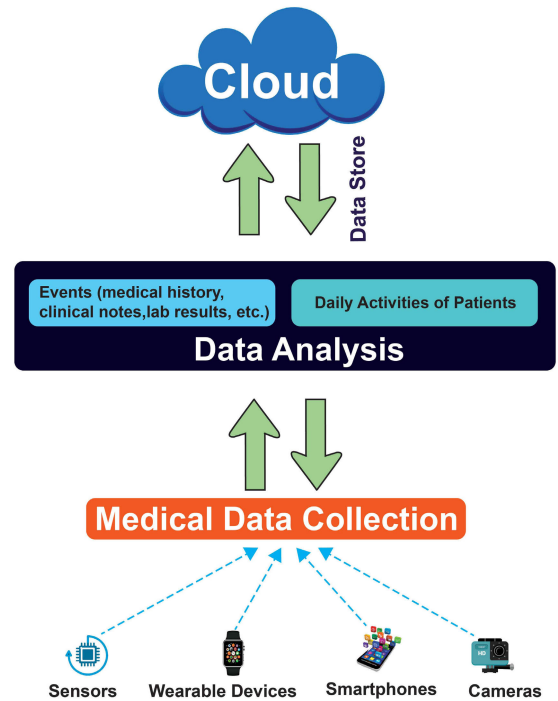

FIGURE 11. Data collection layers in healthcare.

A doctor can track the progress of an elderly patient by using IoT devices that record every detail and activity. Staff members and nurses who care for elderly patients receive information. H-IoT integrates healthcare devices at home for disabilities. Therefore, the doctor may follow and monitor patients' cases at times. Despite this, the data generated and collected are susceptible to various threats. As a result of the attacks, hackers can gain complete command and control (C&C), disrupting the devices [1]. These attacks can drain sensors' power, causing them to fail or run afoul of the entire IoT network in the healthcare industry. Medical data collection includes collecting, analyzing, and monitoring data, as shown in Figure 11.

As a starting point, several types of devices are used for collecting data within the medical data collection layer. These devices include sensors, cell phones, wearable devices, and cameras, among other things. The second phase involves data analysis, which analyzes every event and activity of an elderly patient. A secure server then processes and stores the data. Monitoring healthcare data remotely and automatically recording it includes a patient's medical history and status [201].

### B. MEDICAL APPLICATION LAYER
The traditional application layer in IoT environments handles data processing, storage, and user interfaces. On the other hand, the medical application layer is uniquely designed for handling healthcare-related applications, data, and services, ensuring it follows the highest regulations following the EU General Data Protection Regulation (GDPR) or Health

**TABLE 2.** Comparative analysis of techniques and targeted attacks in published research.

| Reference | Techniques Used | Targeted Attacks | Key Benefits | Limitations | Dataset Employed |
|---|---|---|---|---|---|
| Wagan et al. [193] (2023) | Dynamic Fuzzy C-Means clustering, Customized Bi-LSTM technique. | Botnet-based DDoS, Zero-day network attacks. | - Separates attacks from regular IoMT data.<br>- Securely handles medical data.<br>- Analyzes attack patterns in IoMT. | - Inequity in the model.<br>- Model decisions are not comprehensible.<br>- Possibility of corruption of data. | WUSTL-EHMS-2020 is a heart disease dataset with 36 attributes and 18940 instances developed with the Enhanced Healthcare Monitoring System (EHMS). |
| Thulasi et al. [194] (2023) | Multi Step Convolutional neural network Stacked Long short term memory architecture (MSCSL), light spectrum optimizer (LSO) for optimizing MSCSL's hyperparameters. | Port scans, Brute force, DoS attacks. | - Data adaptability improved.<br>- Reduced computational complexity.<br>- High precision and accuracy. | - Privacy issues are not addressed.<br>- Not optimized for multi-cloud/fog-based dynamic. environments with multiple devices. | Supervisory Control and Data Acquisition International Electrotechnical Commission (SCADA IEC 60870-5-104) datasets. |
| Gupta et al. [195] (2022) | Tree classifier-based network intrusion detection model. | Man in the Middle Attack (Data Alteration and Data Spoofing). | - Reduced input data dimension effectively.<br>- Accelerates anomaly detection. | - The dataset only consists of data alterations and data spoofing attacks.<br>- Tested on a restricted network.<br>- Difficulties in formulating IoMT security design. | A combined network flow and biometrics dataset by [196] |
| Saif et al. [197] (2022) | Hybrid Intelligent Intrusion Detection System (HIIDS), known-nearest Neighbor (kNN) and decision tree (DT), Hybrid approach for feature selection and classification. | Cyber-attacks such as Ipsweep, Portsweep, Smurf, and Brute Force target cloud servers containing Electronic Health Records (EHR). | - Malicious traffic detection and prevention in H-IoT applications.<br>- Metaheuristic algorithms for feature selection reduced computation costs.<br>- Hybrid approach improves intrusion detection accuracy. | - Uncertain performance under high-traffic or larger datasets. - The comparison is limited to Decision Tree and kNN models. There are several ML algorithms, and some are preferable for particular types of data or threats than others. | The NSL-kDD dataset contains 41 features with 125,973 samples. |
| Ray et al. [198] (2022) | A novel DDoS detection algorithm, DDoS preventive algorithms. | Distributed Denial of Services (DDoS) attack. | - Effectively detect DDoS attacks.<br>- A system that restricts access to attackers. | - Other DDoS attacks may not be suitable for this approach.<br>- The system may become complex if nodes do not filter false data. | Cloud-based simulated dataset. |
| Kumar et al. [199] (2022) | A fuzzy min-max neural network (FMMNN) based on fuzzy sets, Supervised algorithm for classification. | DoS, Remote to Local (R2L), User to Root (U2R). | - Nonlinear class boundaries can be learned in one pass.<br>- Detects intrusions more effectively.<br>- Reduced error rate. | - The training time of the FMMNN algorithm needs to be optimized. | NSL-KDD dataset. |
| RM et al. [200] (2020) | ML algorithms (Supervised & unsupervised), Deep Neural Network (DNN). | Replay, Man-in-the-middle, Impersonation, Privileged-insider, Remote hijacking, Password guessing, DoS attacks, Malware attacks. | - Classification & detection (network/host).<br>- Detects anomalies reliably.<br>- DNN compared to other ML algorithms. | - Unpredictable attacks.<br>- Scalability issues.<br>- Network behavior changes rapidly.<br>- Evolving attacks quickly. | Kaggle benchmark dataset for intrusion detection. |

Insurance Portability and Accountability Act (HIPAA) [86], [202] while processing United States healthcare data. This layer presents unique challenges and complexities. For example, electronic health records (EHRs) for patient data, remote consultation platforms, real-time monitoring systems, medical resources, medical care, and diagnosis personnel.

Moreover, the medical application layer processes organizes, and maintains medical records. This layer ensures that patients' data is authentic, secure, private, and reliable when transmitted over the communication system [203]. Numerous application layer protocols exist, including Message Queuing Telemetry Transport (MQTT), Hypertext Transfer Protocol (HTPP), WebSockets, RESTful, Secure-MQTT, and Constrained Application Protocol (CoAP) [204]. Choosing the appropriate protocol for H-IoT depends on its application [205].

### C. ROUTING ATTACK AND NETWORK LAYER

The network layer defines paths for the routing of data over the network. It has been demonstrated that low-power wireless networks with multiple hops are susceptible to a wide range of attacks, with routing attacks emerging as one of the most significant threats in the H-IoT environment [206]. There are many routing attacks, including selective forwarding attacks and replay attacks [207]. In the selective forwarding attack, control packets are deliberately forwarded within the Routing Protocol for Low-Power and Lossy Networks (RPL) while data packets are dropped. In conjunction with sinkhole attacks, this strategy can disrupt established routing paths and lead to severe consequences for the network [208]. A replay attack entails capturing and re-transmitting packets captured from nearby nodes by an unauthorized node or attacker [209]. The purpose of these attacks is to manipulate or obstruct the transmission of data packets to impair the integrity of the network [210]. Furthermore, rank and wormhole attacks are susceptible to IoT routing protocols due to their lightweight nature and limited computational resources. Using these attacks, IoT infrastructures can be devastated by attacks targeting control messages and resources [211].

Routing and network layers receive data sent by medical data collection layers. These layers use protocols such as Wi-Fi 6, Narrowband Internet of Things (NB-IoT), Bluetooth, IPv6 Over Low-Power Wireless Personal Area Networks (6LoWPAN), RPL, WiMAX, ZigBee, Sigfox, LoRa, and 5G NR (New Radio) to transmit data to the medical application layer [212], [213], [214]. This setup requires the analysis of routing attacks, potentially compromising confidentiality and accessibility of medical data. To ensure reliable and secure data transmission, protocol configurations designed for power-efficient and unstable networks should be carefully reviewed for vulnerability to various routing attacks [215]. Following the discussion of routing attacks and network layers, this section explores the use of machine learning to detect such cyber attacks in H-IoT environments.

As highlighted by [216], H-IoT networks are vulnerable to routing attacks, such as sinkholes and wormholes. The authors have introduced various ways to detect such attacks. As a result of the limited resources of devices used in H-IoT, intrusion detection systems cannot be used. The proposed method efficiently detected and handled network attacks using machine learning and deep learning while considering device limitations.

According to [217], especially in e-health applications, protecting sensitive patient data from routing attacks and security threats is pivotal in 5G-IoT. Cloud-based e-health data is at risk of various cyber-threats (i.e., ransomware, loss of privacy, and digital identity fraud) and attacks mentioned in section II. CNN-DMA is a deep learning model that detects malware attacks using a Convolutions Neural Network (CNN). The e-health apps enabled with 5G-IoT and deep learning models such as CNN-DMA are useful for ensuring data safety and system security against cyber-attacks. The deep learning models can be trained directly from original data, such as images and text [218]. Therefore, raw data does not need to be preprocessed before being used for training. Moreover, deep learning algorithms enable the seamless execution of security protocols without consuming significant computing resources, which is one of the major benefits of 5G cybersecurity [219].

In light of the delicate nature of healthcare data, enhancing security within H-IoT's network layers is imperative. The examples show that leveraging machine learning and deep learning can neutralize and mitigate these risks. Following this sub-section, the discussion will focus on the COOJA simulator, a useful tool for studying and understanding these attacks in a simulated IoT environment to help develop secure network protocols.

### D. COOJA SIMULATOR

The COOJA simulator runs on Contiki OS, a portable operating system with limited resources designed specifically for devices such as sensor nodes [220]. It is built upon an event-driven kernel while also supporting multi-threading. It supports a complete TCP/IP stack through IP and programming protothreads. Simulating sensor nodes based on their real characteristics is the main advantage of the COOJA simulator. The process executes program code from ContikiOS and TinyOS using Java Native Interface (JNI). A Java Virtual Machine and C programming (a programming language commonly used in firmware sensor nodes) are interconnected by the JNI. As a result, COOJA can provide accurate simulations of sensor nodes or devices across multiple platforms, including H-IoT, closely replicating real-world functionality. Moreover, it is crucial to note that the primary objectives of COOJA are extensibility and plug-in support. In contrast to the interface, the plug-in allows users to interact with the simulator. Plug-ins, for example, enable users to control simulation speeds or observe network traffic between nodes [221]. COOJA also has several inherent tools, including the "radio logger," which records all packets dispatched by nodes in the simulation and associates them with a universal timestamp [30]. As a versatile simulation tool, COOJA allows for research across diverse environments by generating and collecting data to thoroughly assess wireless sensor networks and body area networks (WBANs) in the H-IoT domain. Using emulated H-IoT devices, the COOJA simulation can even construct attack scenarios to uncover cybersecurity vulnerabilities within H-IoT networks.

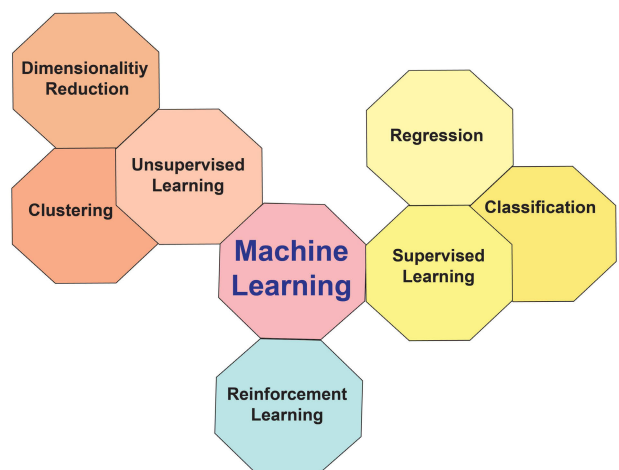

**FIGURE 12.** Machine learning techniques for H-IoT cybersecurity.

Hence, machine learning and deep learning techniques have been explored to prevent routing attacks on H-IoT networks. By reviewing examples, it becomes apparent that artificial intelligence models significantly defend against routing attacks in 5G-IoT e-health. COOJA has also been demonstrated to collect data from H-IoT devices.

## VI. MACHINE LEARNING-DRIVEN INTRUSION AND EVENT DETECTION IN HEALTHCARE-IoT

To ensure data security within the healthcare Internet of Things, it is necessary to explore a variety of strategic mechanisms. The section endeavors to illustrate the pivotal role of machine learning in establishing and reinforcing robust security protocols in healthcare, generalized ML techniques, and exploring potential cybersecurity applications. Following this, the discussion discusses how feature extraction is essential to identifying threats accurately and mitigating risks. Afterward, intrusion and event detection systems designed for healthcare environments will be offered, emphasizing mechanisms for identifying potential breaches. A discourse on the reliability of H-IoT systems is required to achieve consistent and reliable risk mitigation. The last topic of this section is the potential and challenges associated with convolutional neural networks (CNNs) in the context of IoT security.

### A. MACHINE LEARNING TECHNIQUES IN H-IoT CYBERSECURITY

The machine learning method involves using data and algorithms to create models replicating human learning processes. ML algorithms can be refined and optimized by improving an algorithm's loss function [222]. As depicted in Figure 12, ML is traditionally classified into three categories: supervised, unsupervised, and reinforcement learning [223], [224].

Over the past several years, machine learning techniques have been successfully applied to a diverse spectrum of cybersecurity challenges [225]. In the specialized domain of H-IoT cybersecurity, these techniques have emerged as key assets, significantly enhancing the security level of healthcare devices and systems [226]. As illustrated in Figure 13, several potential use cases are encompassed, including intrusion detection and prevention system, classification of H-IoT devices, anomaly detection and prevention, H-IoT attacks classification, zero-day attack detection, predictive analytics for threat anticipation, identity, and access management (IAM), data breach prediction, Cloud anomaly detection, as well as H-IoT devices behavior analysis. According to Table 2, defenders can identify, evaluate, and mitigate possible attacks more precisely with machine learning, as outlined in section IV. Due to this, machine learning algorithms can automate various tasks, including identifying vulnerabilities, deceiving, and disrupting attacks [227]. This section discusses several potential use cases for ML in H-IoT cybersecurity.

- **Classifying Anomalies in Cybersecurity:** In classifying threats, machine learning offers a swift and efficient method for processing huge amounts of data. Machine learning analyzes behavior and continuously changes parameters to find anomalies that could indicate attacks [228]. Machine learning allows for intelligent security services learning that detects or predicts cyber attacks and anomalies.
- **Predicting and Responding to Data Breaches and Cyber Attacks in Real-time:** With machine learning, large amounts of data can be analyzed from multiple sources to predict cyber threats [229]. A machine learning system can quickly build defensive patches in response to an attack after identifying a cyber threat without human intervention [230].
- **Automating processes:** In the H-IoT cybersecurity sector, machine learning is becoming increasingly crucial as time-consuming tasks such as vulnerability assessments, malware analysis, network log scrutiny, and intelligence assessments continue to multiply [231]. Automating security workflows enables organizations to respond quickly to threats and incidents, a feat that cannot be achieved through manual efforts alone [232]. Furthermore, automation offers sustainability by enhancing efficiency and scalability features and reducing unnecessary costs.
- **Authentication and Access Control:** Authentication technology is essential for medical devices since it validates a user's credentials with those stored in authorized user systems or authentication servers, enabling access to multiple systems simultaneously [233]. Machine learning determines when to request multi-factor authentication with intelligent authentication.

### B. FEATURE EXTRACTION
H-IoT feature extraction involves selecting and transforming relevant data attributes. Moreover, it is necessary to determine the most crucial features before embarking on any ML design process [234]. Specific attributes can reduce computing costs and improve storage efficiency [235]. For feature selection, we use the following methods:
- **Filter Methods:** This method calculates the correlation between attributes by comparing the results of their correlations. We select attributes based on statistical methods scores, based on the scores and the threshold values. There are a few popular methods for extracting features from data, such as correlation, information gain, and Chi-Square tests [236].
- **Wrapper Methods:** An ML model trains on selected features. The model's accuracy determines whether to include or exclude attributes depending on its reliability within the subset. Eliminating backward and selecting forwards are the most common methods [237].
- **Embedded Methods:** This method combines the benefits of both filters and wrappers (i.e., extracts the best features while maintaining the computational cost). These methods include Least Absolute Shrinkage and Selection Operator (LASSO) regularization, Random Forest [238].

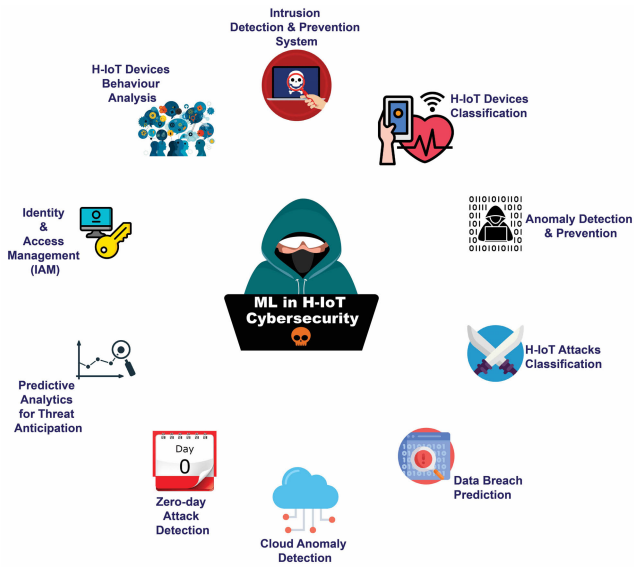

**FIGURE 13.** ML use cases for H-IoT cybersecurity.

In addition, this stage calculates parameter weights for unique features based on the training set using a deep learning algorithm. Each deep learning algorithm has several invisible layers embedded within the inputs and outputs, allowing for the extraction of detailed features [239].

### C. INTRUSION DETECTION SYSTEM IN HEALTHCARE-IoT
Intrusion detection systems for H-IoT are described in this section [240]. An intrusion typically targets a network or device's integrity, availability, or confidentiality by impairing its security. Due to the sensitive nature of healthcare data, H-IoT is a lucrative target for external or internal attackers. Monitoring and analyzing device and network activities is automated by IDS solutions, which are available as hardware or software solutions. IDS consists of three components: information source, analysis, and response. Information sources feed data to the analysis component, which, upon identifying an attack, triggers a response - either passively, such as notifications, or actively, such as disabling communication.

In [241], the authors propose a 5G-driven healthcare landscape where the Internet of Medical Things enables remote patient monitoring. Nevertheless, ensuring the security of data remains a challenge. A new lightweight Intrusion Detection System for IoMT, utilizing kernel techniques for feature selection and a kernel extreme learning machine for decision-making, is introduced. This IDS detected anomalies with 99.9% accuracy on the dataset WUSTL-EHMS-2020.

In [242], the authors used mobile agents to detect intrusions in a medical environment. Furthermore, a simulation of hospital network topology was conducted to simulate IoMT experiments, which included the identification of network-level intrusions using machine learning and regression algorithms and the implementation of various digital imaging and communications in medicine (DICOM) protocols by connected devices such as ultrasound scanners and MRI machines, utilizing wireless body area networks. These networks are composed of wearable and implantable devices that transmit physiological data continuously, allowing patients to be monitored, diagnosed, and treated continuously. In total, 72 independent simulations were conducted across 216 network types, resulting in an overall best- and worst-case detection accuracy of 99.9% and 92.91%, respectively.

In [243], the proposed system integrates ensemble learning and cloud-based architecture to detect cyberattacks. In this ensemble setup, Decision Trees, Naive Bayes, and Random Forests are used as individual learners at the initial level. In the following level, XGBoost uses the classification results to differentiate between normal and attack cases. Using large-scale, diverse IoT networks, ToN-IoT provides an accurate dataset for the model. Experimental results indicate that the proposed framework can achieve a detection rate of 99.98%, a precision of 96.98%, and a reduction in false alarms.

By demonstrating significant accuracy and reduced false alarms in cybersecurity, these examples show that machine learning enhances intrusion detection systems in H-IoT.

### D. EVENT DETECTION SYSTEM
A H-IoT system uses event detection to identify and capture significant occurrences and incidents. An event detection algorithm analyzes sensor data, network traffic, and device interactions to detect patterns, anomalies, and predefined triggers that indicate critical events, such as emergencies, device malfunctions, and abnormal patient conditions. Several machine learning techniques investigate classification, including supervised, unsupervised, and reinforcement learning. A supervised learning algorithm employs a labeled dataset for training that is used to determine the relationship between inputs and outputs. Based on the correlation between the input samples, unsupervised learning algorithms can classify the provided data into clusters. The third category of algorithms is reinforcement learning and online learning. These algorithms utilize current knowledge and explore the environment to classify the data [244], [245].

### E. RELIABILITY IN HEALTHCARE-IoT SYSTEMS
Data exchange between e-health systems must be reliable and efficient in H-IoT for enabling cyber resilience, security metrics (Confidentiality, Integrity, Availability (CIA) Triad), and controls within the ecosystem [1], [137], [246]:
- **Confidentiality:** It is illegal for unauthorized individuals to access medical information.
- **Integrity:** There is no possibility of adversaries altering data during transit or storage.
- **Non-repudiation:** Data transmission and reception are inevitable.
- **Data Freshness:** There is no way to re-generate old data.
- **Resilience to Attacks:** The system should adapt to failed nodes, and there should be no single point of failure.
- **Data Authentication:** It is necessary to authenticate the addresses and object information retrieved.
- **Access Control:** Providing access control to data is a requirement for information providers.
- **Client Privacy:** A lookup system should only be able to infer its purpose from the information provider.
- **Fault Tolerance and Self-healing:** An IoT device or battery energy failure should not disrupt health service continuity.

### F. CONVOLUTIONAL NEURAL NETWORKS (CNNS) IN HEALTHCARE-IoT
Using Convolutional Neural Networks has been demonstrated to be significantly more effective than previous methods in identifying threats, surpassing techniques like Decision Trees, SVMs, and K-nearest neighbors (KNNs) [247]. In classification problems, CNNs are used as feed-forward networks [248]. Essentially, Convolutional Neural Networks are characterized by two processes. First, convolution involves transforming inputs into outputs using filters or kernels. In particular, CNNs have streamlined image recognition to self-contained, removing the need for external image processing software. Furthermore, it can recognize outcomes efficiently and adapt seamlessly to changing identification criteria. Moreover, their ability to leverage preexisting networks enhances system versatility and applicability [249]. Neurons are arranged in layers in a three-dimensional configuration, spanning width, height, and depth. As shown in Figure 14, which illustrates a CNN architecture for analyzing H-IoT data.

As part of the architecture of Convolutional Neural Networks, the initial layer, called Layer 1, is composed of a significant number of neurons to effectively process input data [250]. There are three fundamental parameters: dataset size, quality, and type. Multiple receptive layers can be used to process input layers effectively. This configuration creates overlapping input regions, allowing high-resolution outputs from the original input. For detecting features, CNNs employ operations such as pooling and convolution. Using the fully connected layer as a classifier follows extracting features. Typically, the architecture of CNN, convolutional, and pooling layers are merged, though it is not mandatory.

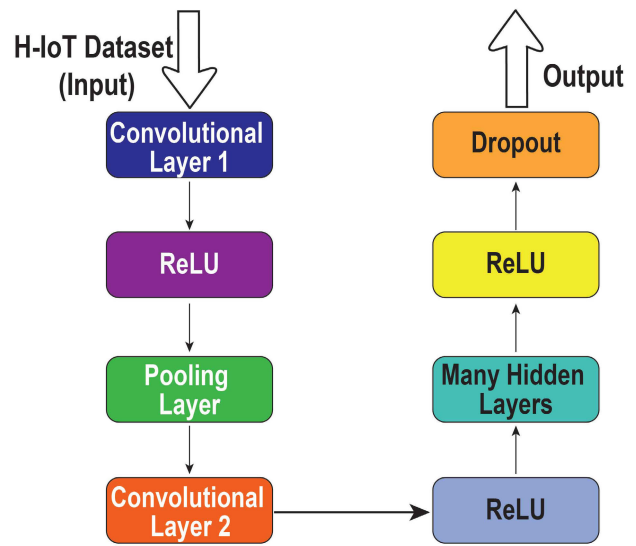

**FIGURE 14.** The convolutional neural network architecture.

#### 1) RECTIFIED LINEAR UNIT (RELU)
In neural networks, ReLUs follow convolution modules. The primary function of ReLU is to activate specific neurons within the network. The Rectified Linear Unit (ReLU) keeps the values of active neurons while converting inactive neurons to $0$ using a simple threshold [251], [252]. The activation information retained by this segment of the neural network remains significant for future inferences made by other network segments.

An activation function like the ReLU is essential for introducing nonlinear operations into inherently linear neural networks. Adding nonlinearity to the network increases its ability to fit data to capture intricate patterns and relationships accurately.

The ReLU activation function does not alter the dimensions of the input or output data. A mathematical expression for this function can be found in equation (1)

$$y = \max\{0, x\} \tag{1}$$

#### 2) POOLING LAYER
Pooling layers contribute significantly to CNN performance by strategically inserting them between CNN layers. By reducing the number of network parameters and down-sampling the data, this integration achieves two objectives:

increasing computational efficiency and reducing computation time. Furthermore, the downsampling process, which is more precise than average pooling, preserves texture information by minimizing the differences in evaluation values resulting from the convolutional layers [253], [254]. Max pooling is essential because it detects and retains the maximum value within each receptive field. Consequently, it preserves essential features while reducing data dimensions, and in each window, it displays the maximum value [255].

### 3) FULLY CONNECTED LAYER

The Fully Connected Layer follows the pooling and convolution layers. Based on the features extracted by the previous layers, this layer classifies the input data. The activation functions of the neurons in this layer perform a significant role in processing the information from the preceding layer. Fully connected neurons allow the network to recognize complex patterns and relations. According to [217], CNN deep learning classifiers can be used to identify malware instances in healthcare datasets. This research demonstrates the networks' capability to analyze intricate data and provide valuable insight into several domains vital to human health and security.

### 4) SOFTMAX

The Softmax activation functions are commonly assigned to the neural networks' final layer to perform multi-class classification. Data can be transformed using the Softmax function into a probability distribution from *0* to *1* with a sum of *1*. As a result of the function, data classification is easier since it emphasizes differences between input values. According to Sharma et al., [256], a CNN-Bidirectional LSTM architecture is proposed for using deep learning in smart healthcare networks for intrusion detection. The model detects DDoS attacks using softmax activation functions by classifying traffic flows as benign or malicious. Faruqui et al. [257] propose SafetyMed, a specialized Intrusion Detection System for securing the Internet of Medical Things. IoMT-specific intrusions are identified with SafetyMed using CNN and LSTM. In the output layer, Softmax activation functions convert raw scores into probabilities ranging from 0 to 1. This enhances the accuracy of the decision-making process. IoMT vulnerabilities are effectively addressed by SafetyMed, which has an average detection accuracy of 97.63%.

The scope of this section encompassed a wide range of machine learning techniques, applications, and advanced CNN-based mechanisms illuminating multifaceted strategies for protecting H-IoT devices.

## VII. SECURITY AND PRIVACY IN H-IoT ENVIRONMENTS

This section emphasizes the importance of effective safeguards and rigorous protocols to protect sensitive data and preserve user confidentiality in H-IoT environments. In-depth analysis explores key challenges, effective strategies, and emerging trends. IoT is being transformed into a network of interconnected devices, where security and privacy are becoming crucial challenges. Therefore, the IoT network is built around anomaly detection techniques. Anomaly detection algorithms detect abnormal patterns immediately in the IoT, preventing serious problems. However, healthcare and the IoT combine to raise the stakes even higher. As a result, machine learning could potentially provide remedies for these challenges. Additionally, a machine learning-based architecture for anomaly detection is described in detail in the section.

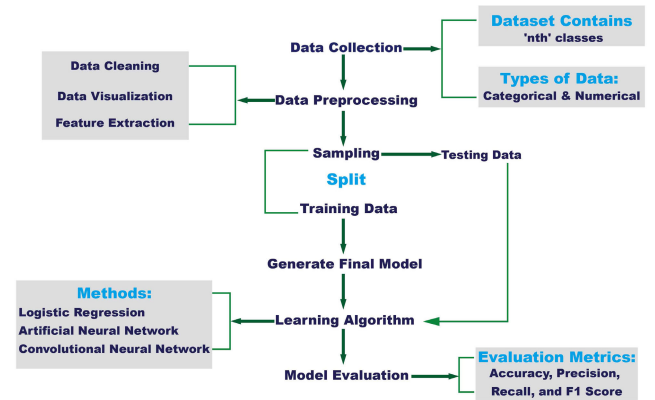

**FIGURE 15.** ML-based anomaly detection architecture in H-IoT.

### A. ANOMALY DETECTION

The term anomaly refers to data segments that do not follow the expected pattern or features of the rest of the data. Anomaly detection is crucial since anomalous activities contain critical information and relevance [258]. Anomaly detection involves several independent processes in H-IoT, as illustrated in Figure 15. Data collection is the first step in this architecture. Following the dataset's creation, it should be analyzed thoroughly to determine the data types.

It is also crucial to preprocess the data before analyzing it [135]. The preprocessing of data includes filtering, visualizing, and extracting features. These steps convert the data into feature vectors. It is then necessary to divide the feature vectors into training and testing sets to an 80:20 ratio. The learning algorithm develops an optimal final model by utilizing the training set. It is possible to use a variety of classifiers to experiment with the most accurate accuracy [259].

With 5G networks facilitating rapid, high-volume data exchanges, the IoMT has gained momentum in the healthcare sector. In [260], the emphasis is on detecting anomalies using deep learning for enhanced security within 5G contexts in IoM. By utilizing multi-model autoencoders for feature extraction, the complexity of traffic feature information has been significantly reduced. A new algorithm for detecting multi-featured sequence anomalies, optimized to cope with the high data volume of 5G networks, is also presented. It has been demonstrated that deep learning combined with 5G technology can identify and mitigate anomalies efficiently, thereby increasing IoMT systems' resilience and reliability.

### B. BEST PRACTICES FOR ENABLING DATA PRIVACY, SECURITY, AND BUILDING RISK MITIGATION STRATEGIES IN SMART ENVIRONMENTS

Digital healthcare environments [261] are becoming more prevalent despite numerous security and privacy concerns, and they are susceptible to a unique set of challenges (i.e., securing the massive amount of healthcare data harvested every second from thousands of connected IoT devices). Healthcare procedures and decisions are made based on this data; in situations where this data is compromised or breached, this may result in a wrong diagnosis/treatment affecting human life. Securing smart environments requires implementing defense-in-depth strategies and multiple layers of security across the ecosystem such as: (i) Standards and data governance risk and control (DGRC) policies for access control, data protection, encryption, remote access (ii) Monitoring and response mechanisms for Security Information and Event Management (SIEM), identity and access management, log analysis, cyber forensics, (iii) Perimeter controls for intrusion detection, prevention and next generation firewalls, (iv) Securing the network using VPN services, web-filtering, securing remote services and site-to-site connections (v) Hardening end-point security by enabling Zero Trust mechanisms, patching the systems with latest updates to mitigate vulnerabilities, using an up-to-date antivirus tool to detect and protect the end-point for new vulnerabilities, (vi) the human element is the weakest link and its essential to build cyber awareness within the workplace environment and train them so that they are cyber prepared, (vii) enabling application security using vulnerability scanners and making sure that only inventoried tools are used, (viii) last but not the least, to ensure data security mechanisms are in place (i.e. end-to-end secured data, data classification and have a data loss prevention/backup strategy) [1]. However, the impact and frequency of cyber attacks have escalated due to generative AI. Since the attacks are constantly evolving and new threats are coming to the surface, there is a possibility that the novel threats would go undetected. This is why there is an essential need for a versatile machine-learning solution to neutralize and mitigate these threats.

### C. MITIGATING HEALTHCARE-IoT SECURITY AND PRIVACY THREATS USING MACHINE LEARNING

The security and privacy of sensitive health information has become a worldwide concern. In healthcare security and privacy, machine learning-based solutions are essential among cryptographic, password-based, and digital signature-based solutions. As shown in Figure 16, machine learning models can detect Device Spoofing and Identity Impersonation, Intrusion Detection and Prevention, and Device Authentication. A machine learning model can prevent anomalies after detection. As a result, ML models perform a significant role in H-IoT [262]. In healthcare systems, Hau et al. applied machine learning techniques to increase security and privacy [263]. Begli et al. implemented SVM-based intrusion detection to monitor remote healthcare [264]. Sengan et al. developed an adaptive, ML-driven routing system for healthcare data security [265].

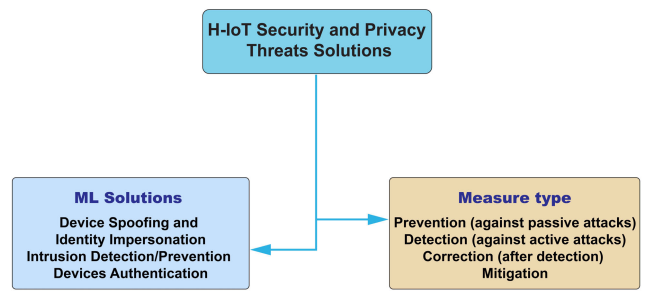

**FIGURE 16.** H-IoT security and privacy threat solution.

## VIII. IMPACT OF HEALTHCARE-IoT TECHNOLOGIES IN THE FUTURE

Healthcare and smart cities benefit greatly from the Internet of Things (IoT). This technology will significantly change everyday life and urban environments in the coming years. This section discusses the potential impact of H-IoT on smart cities, including relevant challenges and innovative solutions. Quality of service is integral to this discussion, which entails ensuring the reliability of services. Furthermore, H-IoT networks can be made more flexible, scalable, resource-efficient, and secure with the help of Software-Defined Networks (SDNs).

### A. FUTURE OF HEALTHCARE-IoT TECHNOLOGIES

Generally, people expect H-IoT systems to integrate with their living environments. The future of infrastructure lies in smart cities. As a result, affiliating the H-IoT system within modern city architecture facilitates widespread deployment of the H-IoT [266]. Software-defined networking plays a significant role in smart cities and H-IoT systems. Using SDN in H-IoT, cities can gain more efficient management, more secure provision of services, effective resource allocation, and overall better performance.

### B. CHALLENGES

Despite the growing popularity of IoT applications in healthcare, several challenges restrict their implementation. In the following section, we explore key challenges connected to H-IoT development [267].

#### 1) HETEROGENEOUS NETWORKS

Developing H-IoT requires devices to interact and interoperate across diverse networks and operating systems. Among the technologies used in communication are Bluetooth, IrDA, UWB, Zigbee, WLANs, Wi-Fi 6, and 5G. Batteries drain out faster when communicating over long distances [268], [269]. It is essential that these devices, regardless of their heterogeneity in implementation and communication, can identify and communicate with each other.

### 2) QUALITY OF SERVICE (QOS)

Healthcare IoT is characterized by low latency and low power consumption. There is a risk to a patient's health when dealing with issues related to the Internet of Things. Consequently, H-IoT signals should be transmitted and processed with the least possible delay. High-bandwidth networking resources can achieve this [270]. A 5G network, with its ability to ensure quality of service through rapid data transmission, minimal latency, and extensive connectivity, presents itself as an ideal solution for H-IoT applications [271]. Quality of service refers to the ability of a system to maintain a basic level of performance and reliability, ensuring that data is transmitted and received accurately and promptly.

### 3) PRIVACY AND SECURITY

In H-IoT, the privacy and security of patients and their data are paramount as any compromise could result in severe consequences (i.e., loss of command and control of a cobot arm used for precision surgeries may lead to its malfunctioning causing harms to person, property, loss of availability could lead to healthcare service disruption leaving critical patients who may need immediate attention vulnerable, loss of integrity would risk patients to wrong diagnosis or medication, loss of confidentiality would lead to loss of privacy, digital identity frauds, deep fakes, bio-hacking issues, etc.). In many cases, H-IoT uses heterogeneous networks to communicate and store data, exposing it to attack over long distances [272]. Thus, the security and privacy of data require anomaly detection, mitigation, robust encryption, and authentication techniques, as mentioned in section VII-B. However, in designing and implementing the defense-in-depth techniques, it is necessary to consider the resource constraints of the devices involved in H-IoT, and the developed solutions should be lightweight and energy-efficient.

### C. SOFTWARE-DEFINED NETWORK (SDN) IN HEALTHCARE-IOT

An emerging paradigm in computer networking is software-defined networking. Data planes, control planes, and application planes are the three tiers of SDN. Data and control planes are separated to improve network performance. In contrast, traditional architectures combine data and control planes onto one device, while SDN separates them to simplify network management [273], [274]. SDNs forward packets according to predefined policies [275]. SDN-based H-IoT consists of Sensor Controllers, Cloud Controllers, Network Controllers, and Closed-Loop Controllers, which provide four types of services: data acquisition, network

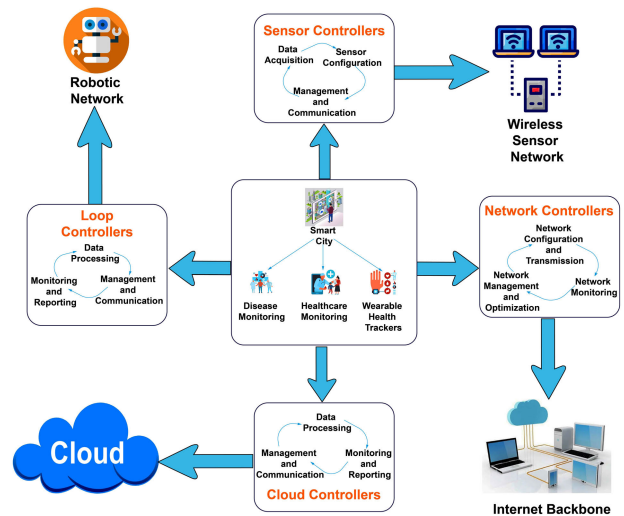

**FIGURE 17.** Software defined network-based H-IoT.

configuration and transmission, processing, and actionable feedback, as illustrated in Figure 17.

The software-defined network is used by Google, for example, to manage its wide-area network traffic [276]. The network infrastructure increasingly relies on SDNs, and protecting a network from attack is crucial to ensure its integrity and availability [277]. In [278], a novel software-defined network architecture is presented, incorporating a new controller and load-balancing algorithms based on machine learning for the healthcare IoT. A support vector machine algorithm, which achieved 95.1% accuracy through extensive simulations, validates the architecture's effectiveness. In preparation for broader 5G integration, this approach represents an important advance in H-IoT. Adding deep learning to SDNs makes H-IoT services, such as telemedicine, more reliable and secure [279]. A software-defined network system illustrates the communication process between medical services and nano-networks inside individuals' bodies [280]. A certain set of software rules controls this method, which makes healthcare more efficient.

## IX. CONCLUSION

This paper systematically dives into digital healthcare's fundamental elements and practical applications. It discusses the essential technologies and methodologies for optimizing these systems within Healthcare 5.0. The research focuses on various cyber threats, vulnerabilities, and cyber attacks carried out at the different IoT layers and explores the security and privacy of H-IoT through deep neural networks and AI. A demonstration of a cloud-based solution and big data analytics using IoT and software-defined networks is provided. This illustrates the advantages of a precise, efficient, and secure healthcare ecosystem. The authors also present insights on the security challenges related to wireless and communication technologies (i.e., 5G) in H-IoT and how

ML algorithms can enable the digital healthcare environment to bridge those security gaps.

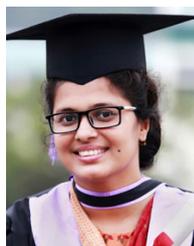

**MIRZA AKHI KHATUN** (Member, IEEE) received the B.Sc. degree in computer science and engineering and the M.Sc. degree in computer science and engineering from Jagannath University, Bangladesh, focusing on Internet of Things (IoT) security and machine learning, in 2019. She is currently pursuing the Ph.D. degree with the Department of Electronic and Computer Engineering, University of Limerick, Ireland. Her ongoing research delves into the healthcare-IoT security and utilizing AI/ML innovations. She has an exemplary academic background, underscored by a scholarship from the Science Foundation Ireland Centre for Research Training in Foundations of Data Science awarded, in 2022. During the master's program, she secured the first position in her class. She has also published research papers at renowned IEEE international conferences. Her research interests include the IoT, artificial intelligence, machine learning, data analysis, network security, and systems architecture. She is also reflecting her deep commitment to the professional community.

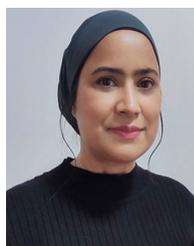

**SANOBER FARHEEN MEMON** (Member, IEEE) received the B.Eng. degree in electronic engineering and the M.Eng. degree in electronic systems engineering from the Mehran University of Engineering & Technology (MUET), Jamshoro, Pakistan, in 2010 and 2014, respectively, and the Ph.D. degree in electronic and computer engineering from the University of Limerick, Ireland, in 2023. She is currently a Researcher on the OXI-SMART Project with the Optical Fibre Sensors Research Centre, Department of Electronic and Computer Engineering (ECE), UL. During the Ph.D. research, she developed plastic optical fiber sensors for in-situ and real-time ultra-low-level ethanol concentration measurement in microalgal bioethanol production applications. She was with MUET, Pakistan, for more than three years as a Faculty Member (a Lecturer and a Visiting Lecturer) and taught a variety of modules to the B.Eng. students. Her current research interests include optical fiber sensors, medical sensors, photonics engineering, instrumentation and measurement, and intelligent systems. She has received multiple funding grants, including the European Union EM INTACT Doctorate Scholarship, UL Postgraduate Residential Scholarship, UL Funding & ECE Departmental Funding, Merit Scholarship from MUET, and some international travel/conference participation grants. She is a member of IEEE WIE, IEEE Photonics Society, IEEE Instrumentation & Measurement Society, and SPIE.

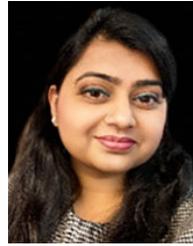

**LUBNA LUXMI DHIRANI** (Senior Member, IEEE) received the B.Eng. degree in computer systems, the M.Sc. degree in business information technology, and the Ph.D. degree in hybrid cloud computing QoS and SLA-based issues in heterogenous cloud environment. She is currently an Assistant Professor with Department of Electronic and Computer Engineering, University of Limerick. She is also the Course Director of the B.Sc. Cybersecurity Practitioner Apprenticeship Program. She is also the first WIE Ambassador from Ireland in the IEEE WIE (U.K.&I Section) and a member of the National Standards Authority of Ireland (NSAI). She has worked on interdisciplinary, industry-based projects securing machine-to-machine communications in industry 4.0, V2I safety and security, and digital healthcare. Her publications are listed at: Google Scholar, she has delivered more than 35 technical workshops, tutorials, masterclasses, panel, and keynote speeches at international conferences and platforms. Her research interests include cybersecurity, privacy, data governance, risk, cyber law, regulations, the IIoT, IT/OT, and cloud and security standards.

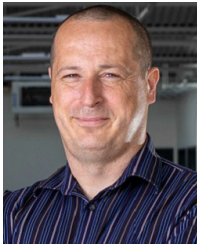

**CIARÁN EISING** (Senior Member, IEEE) received the B.E. degree in electronic and computer engineering and the Ph.D. degree from NUI Galway, in 2003 and 2010, respectively. From 2009 to 2020, he was a Computer Vision Architect Senior Expert with Valeo. In 2020, he joined the University of Limerick as an Associate Professor in AI.

・・・